# Progress on the SOXS NIR Spectrograph AIT

Fabrizio Vitali[1], Matteo Aliverti[2], Francesco D'Alessio[1], Matteo Genoni[2], Salvatore Scuderi[3], Matteo Munari[4], Luca Oggioni[2], Andrea Scaudo[2], Giorgio Pariani[2], Giancarlo Bellassai[4], Rosario Di Benedetto[4], Eugenio Martinetti[4,] Antonio Miccichè[4], Gaetano Nicotra[4], Giovanni Occhipinti[4], Sergio Campana[2], Pietro Schipani[5], Riccardo Claudi[6], Giulio Capasso[5], Davide Ricci[6], Marco Riva[2], Ricardo Zanmar Sanchez[4], José Antonio Araiza-Durán[7], Iair Arcavi[8], Andrea Baruffolo[6], Federico Battaini[6], Sagi Ben-Ami[9], Anna Brucalassi[7], Rachel Bruch[9], Enrico Cappellaro[6], Mirko Colapietro[5], Rosario Cosentino[10], Paolo D'Avanzo[2], Sergio D'Orsi[5], Massimo Della Valle[5], Avishay Gal-Yam[9], Marcos Hernandez Díaz[10], Ofir Hershko[9], Jari Kotilainen[11], Hanindyo Kuncarayakti[12], Marco Landoni[2], Gianluca Li Causi[13], Laurent Marty[5], Seppo Mattila[12], Hector Pérez Ventura[10], Giuliano Pignata[14], Kalyan Radhakrishnan[6], Michael Rappaport[9], Adam Rubin[15], Bernardo Salasnich[6], Stephen Smartt[16], Maximilian Stritzinger[17], David Young[16]

[1]INAF Osservatorio Astronomico di Roma, [2]INAF - Osservatorio Astronomico di Brera, [3]INAF - Istituto di Astrofisica Spaziale e Fisica Cosmica, [4]INAF - Osservatorio Astrofisico di Catania, [5]INAF - Osservatorio Astronomico di Capodimonte, [6]INAF - Osservatorio Astronomico di Padova, [7]INAF-Osservatorio Astrofisico di Arcetri, [8]Tel Aviv University, [9]Weizmann Institute of Science, [10]INAF - Fundación Galileo Galilei, [11]FINCA - Finnish Centre for Astronomy with ESO, [12]Tuorla Observatory, Department of Physics and Astronomy, University of Turku, [13]INAF - Istituto di Astrofisica e Planetologia Spaziali, [14]Universidad Andres Bello, [15]European Southern Observatory, [16]Queen's University Belfast, School of Mathematics and Physics, [17]Aarhus University

## ABSTRACT

The Son Of X-Shooter (SOXS) is a single object spectrograph, built by an international consortium for the 3.58-m ESO New Technology Telescope at the La Silla Observatory, ranging from 350 to 2000 nm. In this paper, we present the progress in the AIT phase of the Near InfraRed (NIR) arm. We describe the different AIT phases of the cryo, vacuum, opto-mechanics and detector subsystems, that finally converged at the INAF-OAB premises in Merate (Italy), where the NIR spectrograph is currently being assembled and tested, before the final assembly on SOXS.

## 1. INTRODUCTION

The Son Of X-Shooter (SOXS) is a single object spectrograph, built by an international consortium for the 3.58-m ESO New Technology Telescope at the La Silla Observatory [1]. It offers a simultaneous spectral coverage over 350-2000 nm, with two separate spectrographs. In this paper we present the progress in the AIT phase of the Near InfraRed (NIR) cryogenic echelle cross-dispersed spectrograph [2], which consists in a fully criogenic echelle-dispersed spectrograph, working in the range 0.80-2.00 mm. The optical design is based on a 4C echelle and the dispersion is obtained via a main disperser grating and three cross-disperser in double-pass. The NIR spectrum will be dispersed in 15 orders and imaged on an Hawaii H2RG IR array from Teledyne, controlled via the new NGC controller from ESO. The spectrograph will be cooled down via a Leybold M10 system. All the subsystems (cryo, vacuum, opto-mechanics and detector) have been designed, procured and tested by INAF Institutes, and finally converged at the INAF-OAB premises in Merate, where the NIR spectrograph is currently being assembled and tested, before the final assembly on SOXS.

*fabrizio.vitali@inaf.it; phone +39 0694286462.

## 2. THE AIT PROCEDURES OF THE NIR SPECTROGRAPH

### 2.1 Optics

The optics have been tested warm on the dummy bench and are ready to be mounted in the cryostat. Figure 1 shows the detail of the SOXS NIR spectrograph input-end sub-bench mounted and aligned on the dummy bench. The input-end sub-bench holds the entrance baffles, the entrance pupil stop, the NIR common path field lens, the slit plate and motor, and the 45° spectrograph folding mirror (Figure 2).

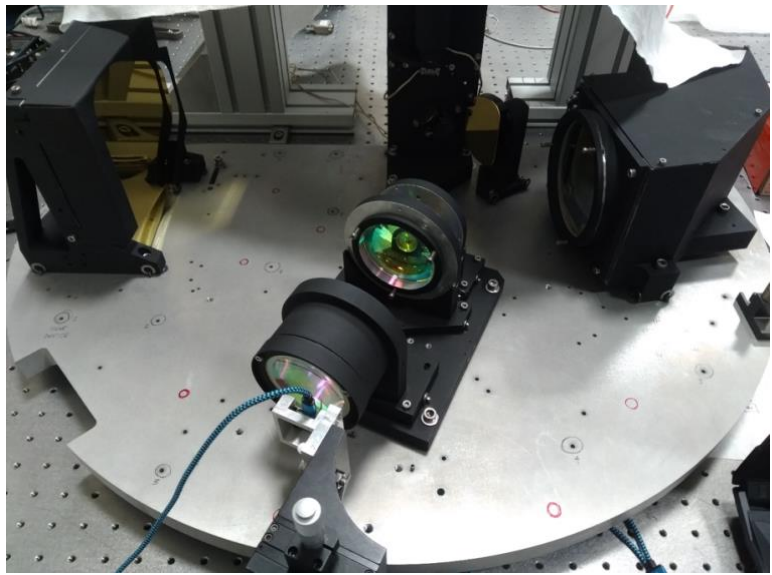

*Figure 1: SOXS NIR spectrograph elements mounted and aligned on the warm dummy bench.*

Being the spectrograph pre-aligned and tested in warm conditions, we used a CMOS 2.2μm pixels to record images of different portions of the diffraction orders 21, 22, 23, 24. Other orders cannot be detected because of the CMOS wavelength cut-off..

The following tests have been done:

- Diffraction orders and wavelength identification: several features of the Ne lamp along the orders have been identified and check the matching with expected from Zemax theoretical model. This also confirms the linear dispersion is within design parameters.

- Inter order separation and order curvature: the above orders have been reconstructed by merging different flat frames (LDLS source). A comparison of the extracted positions along the main dispersion direction is done w.r.t. the Zemax theoretical model. A maximum deviation of the order curvature is about 180 μm, while the inter-order separation deviation from the theoretical model is about 100 μm. An example is given in Figure 3.

- The resolution elements FWHM: tested both using fibers illuminating the common-path simulator and by (quasi-)homogeneously illuminating the common-path simulator pupil stop in order to illuminate the full slit aperture. The FWHM of the analyzed spectral features of Ne Lamp is in agreement with expected values, within 5% deviation in both spectral and spatial directions. Combining this with linear dispersion the resolving power is in agreement with the expected value.

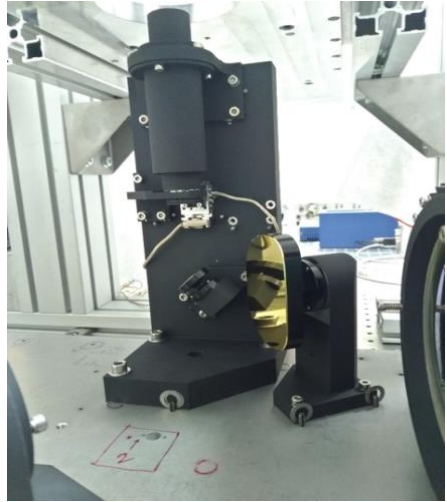

*Figure 2: SOXS NIR spectrograph input-end sub-bench mounted and aligned on the dummy bench.*

An example of the image of an equivalent 0.5" seeing in passing through the 0.5" slit aperture, for wavelength 849.536 nm, is shown in Figure 4; the image is sampled with CMOS 2.2μm (not the real detector pixels of 18μm size). A residual astigmatism can be noticed, resulting in a non-perfectly circular shape; this is caused by the Wave Front Error rms residuals introduced by the reflection grating.

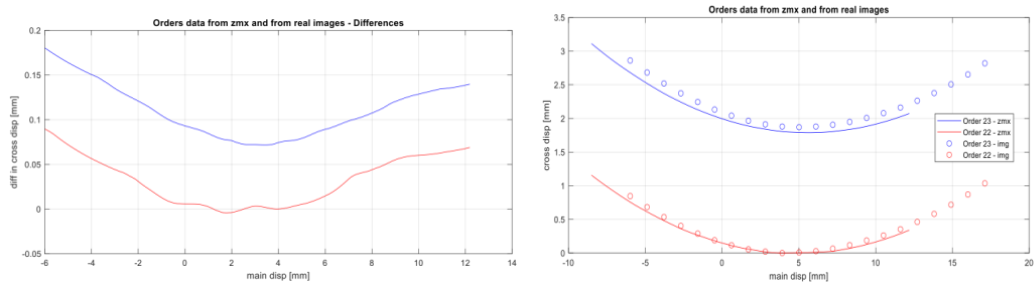

*Figure 3: example of trace for diffraction orders 22 and 23.*

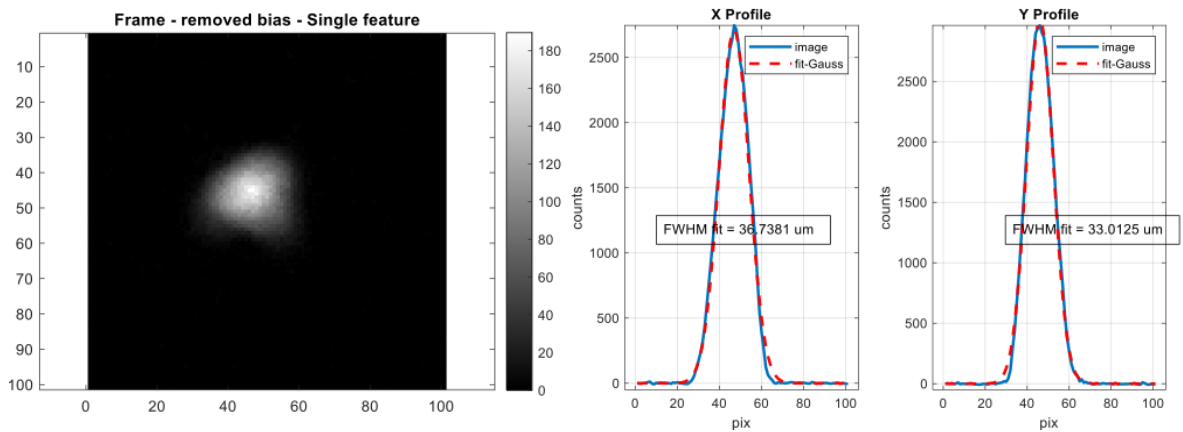

*Figure 4: Left: image of the 849.536 nm feature of the Ne calibration Lamp related to an equivalent 0.5" seeing through the 0.5" slit aperture recorded with the CMOS 2.2μm. Right: main dispersion and cross dispersion binned profiles; these are well fitted with a Gaussian from which the extracted FWHM is in agreement with expected value at this wavelength.*

## 2.2 Mechanics & Criogenics

The NIR cryostat has been initially tested on the main flange of SOXS (Figure 5, left), for a static load test. Then, it was moved on a bench for the mounting of all internal wiring and mechanical dummies (Figure 5, right). All the temperature resistors and heaters have been wired and tested.

The cryostat is currently populated with dummies and is under intense cryo-vacuum tests (Figure 6), with the CryoCooler system working as expected. The details of the mechanical design can be found in [3].

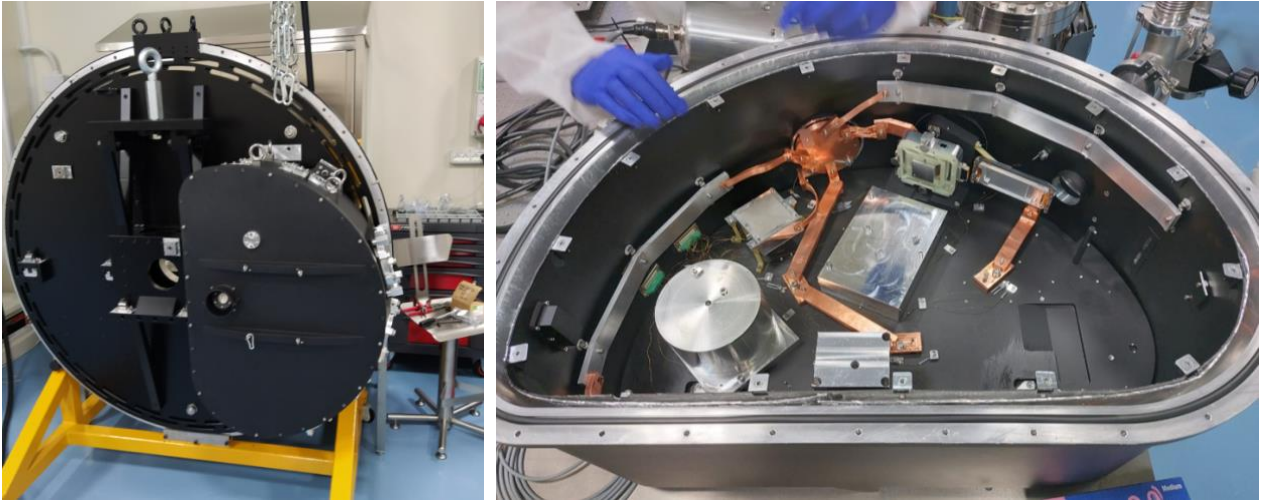

*Figure 5: left: the D-shaped NIR cryostat mounted on the SOXS flange, right: open view of the NIR cryostat: the dummies, the cold straps and the cold head can be clearly seen.*

The cryocooler cold head is thermally linked, via the cold straps, with the thermal shield, the bench, all the opto-mechanics, the cryo-pump and the detector mounting. We can fine tuning the cooling rate of the different part of the bench, modulating the tightening of the screws at the straps junctions.

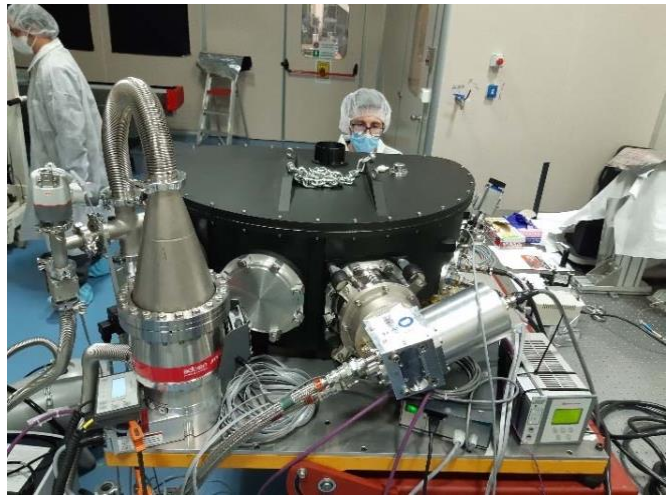

*Figure 6: the NIR cryostat during the cryogenic tests at the bench. In the foreground, you can see the Turbo Molecular Pump and the Cryocooler Head.*

The vacuum and cryogenics system (but the detector) is controlled via a Siemens PLC, in which all the procedures necessary to control the system reside. The first results show that the system can monitor the temperatures changes and the heaters can maintain the temperatures at the desired constant level. Details of the cryogenics system and tests can be found in [4].

**2.3 Detector System**

The detector control system (NGC, LLCU and MUX) was initially mounted and tested in the INAF-OAR laboratory (Figure 7, left), where a "first light" has been obtained, feeding the array with a green laser spot (Figure 7, right).

The Detector system is currently operational on a bench in the INAF-OAB-Merate Clean Room and under testing (Figure 8, left). The system is reachable remotely through the Observatory VPN, to allow for the full NGC operation in remote mode, both from INAF-OAR Observatory and ESO Garching lab to perform tests and mainteinance.

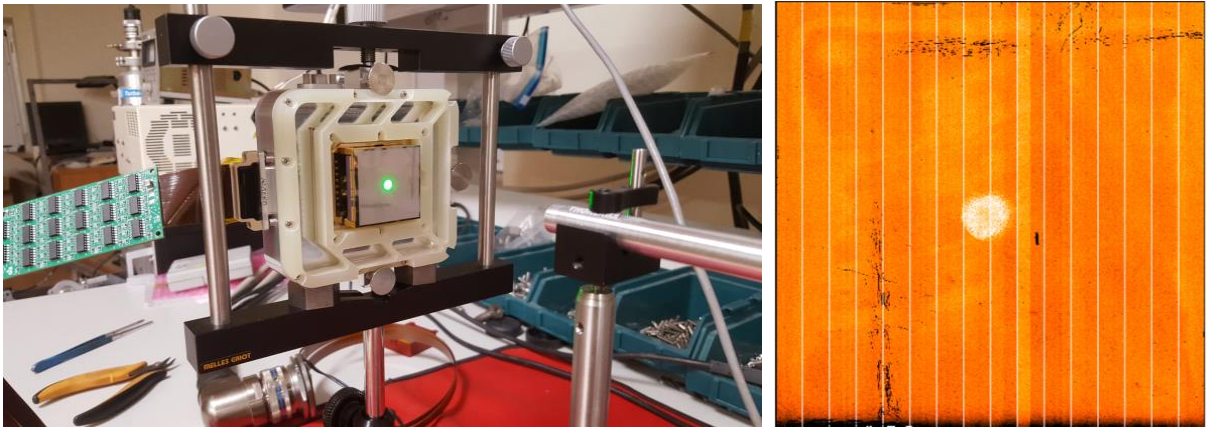

*Figure 7: left: the MUX mounted at bench at INAF-OAR laboratory, for a first functionality check of the detector system (MUX/NGC/LLCU), right: a green laser spot, as imaged by the MUX.*

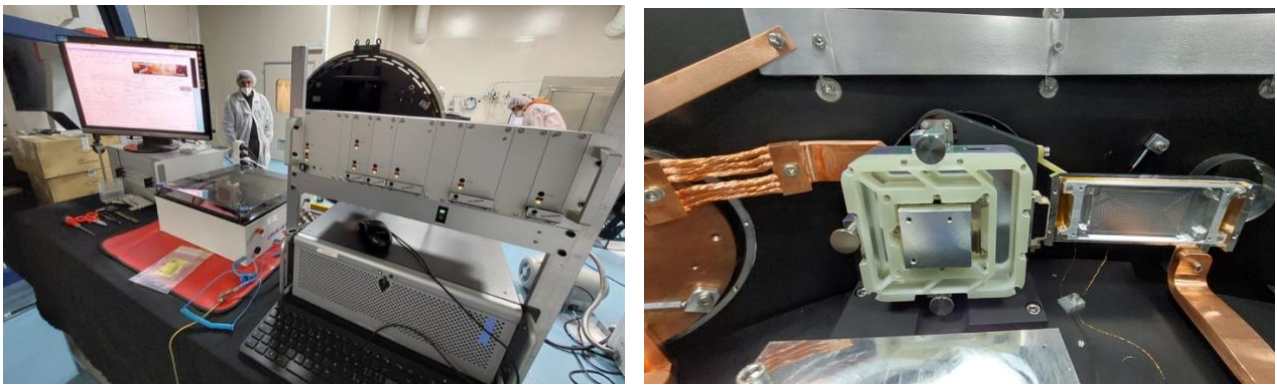

*Figure 8: left: the detector system on the bench in the clean room in INAF-OAB-Merate, right: the array holder mounted in the cryostat.*

The array holder and the annexed cryogenics electronics holder are already mounted in the cryostat, with a mechanical dummy for the cryo-vacuum tests (Figure 8, right). The temperature of the detector is controlled by a Lakeshore 336 temperature controller, with which we can set the final working temperature of the detector (40 K) and the cooling rate, that we initially set at 2 K/min, for the safety of the detector. We have also tried slower rates and we can cooldown the detector as slow as 0.5 K/min, a rate absolutely safe for the H2RG detector. In Figure 9 (left), you can see a plot of a cooling down (red line), with a rate of 1 K/min (blue line): once the cooling is started, the heater is switched on (green line) but, after some time it can not contrast the cooling power of the cold head (yellow line), so the rate increase a bit (< 1.5 K/min), to eventually came back to the set value. We now can set the cooling procedure, in order to avoid the over rate phase and ensure a completely controlled cooling rate for the detector.

Once at the working temperature of 40 K, the Lakeshore controller can stabilize the temperature better than the specs of ± 0.01 K, as shown in Figure 9, right panel.

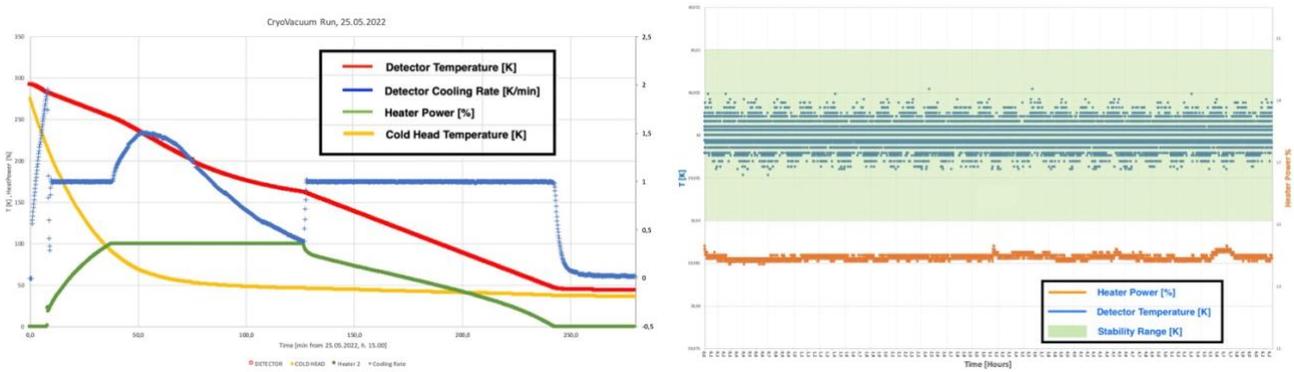

*Figure 9: left: the detector cooling rate @ 1K/min, right: the detector temperature stability @ 40K.*